\documentclass[10pt,conference]{IEEEtran}
\ifCLASSOPTIONcompsoc
  \usepackage[nocompress]{cite}
\else
  \usepackage{cite}
\fi
\ifCLASSINFOpdf
\else
\fi
\usepackage{amsmath,amssymb,amsfonts}
\usepackage{algorithmic}
\usepackage{graphicx}
\usepackage{textcomp}

\usepackage[margin=1in]{geometry} 
\usepackage{amsmath,amsthm,amssymb}
\usepackage[margin=1in]{geometry} 
\usepackage{amsmath,amsthm,amssymb}
\usepackage{bbm}
\usepackage{cuted} 

\usepackage[T1]{fontenc} 
\usepackage[utf8]{inputenc} 
\usepackage{lmodern} 
\usepackage{graphicx}
\usepackage[ruled,vlined]{algorithm2e}

\usepackage[euler]{textgreek} 
\usepackage{slashed}
\usepackage{feynmf}
\usepackage{array}
\usepackage{blindtext}

\usepackage{xcolor}
\usepackage{longtable}

\usepackage{physics} %

\usepackage{hyperref}

\newcommand{\bi}{\text{binary}}

\begin{document}
%
\title{Automatic Generation of an Efficient Less-Than Oracle
for Quantum Amplitude Amplification}

\author{\IEEEauthorblockN {Javier Sanchez-Rivero\IEEEauthorrefmark{1},
Daniel Talaván\IEEEauthorrefmark{1},
Jose Garcia-Alonso\IEEEauthorrefmark{2}, \\
Antonio Ruiz-Cortés\IEEEauthorrefmark{3}, 
and Juan Manuel Murillo\IEEEauthorrefmark{4}} \\
\IEEEauthorblockA{\IEEEauthorrefmark{1}COMPUTAEX Foundation \\ 
\IEEEauthorrefmark{2}University of Extremadura \\ 
\IEEEauthorrefmark{3}University of Seville \\
\IEEEauthorrefmark{4}COMPUTAEX Foundation and University of Extremadura}
}


%


\maketitle

\begin{abstract}
Grover's algorithm is a well-known contribution to quantum computing. It searches one value within an unordered sequence faster than any classical algorithm. A fundamental part of this algorithm is the so-called oracle, a quantum circuit that marks the quantum state corresponding to the desired value. A generalization of it is the oracle for Amplitude Amplification, that marks multiple desired states. In this work we present a classical algorithm that builds a phase-marking oracle for Amplitude Amplification. This oracle performs a less-than operation, marking states representing natural numbers smaller than a given one. Results of both simulations and experiments are shown to prove its functionality. This less-than oracle implementation works on any number of qubits and does not require any ancilla qubits. Regarding depth, the proposed implementation is compared with the one generated by Qiskit automatic method, \textit{UnitaryGate}. We show that the depth of our less-than oracle implementation is always lower. This difference is significant enough for our method to outperform \textit{UnitaryGate} on real quantum hardware.
\end{abstract}


%
\IEEEpeerreviewmaketitle

\section{Introduction}
\label{sec:introduction}

The arrival of the noisy intermediate-scale quantum (NISQ) era \cite{Preskill2018quantumcomputingin} where quantum computers with several qubits (quantum bits) are available \cite{qubits_1,qubits_2,qubits_3} has enabled the possibility of testing the theoretical work on quantum algorithms on real devices \cite{3-qubits_grover, 15_shor}. This fact has enabled a great research effort that is currently being devoted to the optimization of existing algorithms and to the creation of new ones.

Quantum algorithms usually take a fully superposed state as starting point and manipulate it to reach another quantum state where some of the states have increased amplitude. These states are the most probable solution to the problem being addressed. The possibility of having superposition states and the way amplitudes are manipulated means that certain problems can be solved more efficiently by quantum computers than by classical ones \cite{15_shor,variational_quantum}. A well-known example of this is Grover's algorithm. 

Grover presented \cite{grover1} a quantum algorithm that searches in an unordered sequence faster than any known classical algorithm. 
 Grover's algorithm works by combining two quantum operations. The first one marks (by giving it a $\pi$-phase) the desired quantum state. This marking operation is known as oracle.  The second operation aims to amplify the amplitude of the marked state. This operation is called diffusion operator or diffuser. In order to reach such amplification it is often needed to repeat the pair oracle-diffuser several times. The generalisation of Grover to amplify multiple states is Amplitude Amplification (AA) \cite{AA, Grover_1998}.


In this paper we present the implementation of an oracle that performs a less-than operation. Taking as input a quantum state in which qubits encode natural numbers (including 0), this oracle gives a π-phase to those numbers less than a given one. In addition, a classical algorithm is provided that automatically generates the quantum circuit corresponding to the oracle for any number of qubits.

 The complexity function of this less-than operation is $\mathcal{O}(\sqrt{N/M})$, where $N$ is the number of naturals encoded in the quantum state and $M$ ($1\leq M\leq N$) is the number of naturals less than the given number. Thus, the complexity function is bounded above by $\mathcal{O}(\sqrt{N})$ (better than classical $\mathcal{O}(N)$ for an unordered sequence of naturals). Moreover, the provided implementation avoids the use of ancilla qubits and produces circuits with a depth low enough to be realizable in real quantum hardware.


The result presented here is part of a research work that aims to achieve reusable and composable quantum operations. For example, from the less-than a greater-than oracle can be easily obtained, and by combining both, ranges of integers can also be produced.

The rest of the paper is organised as follows. Next, in Section \ref{sec:background} the background of this work is provided. Then, a detailed description of the classical algorithm for the automatic generation of the less-than implementation is presented in Section \ref{sec:classical}. 
Results of simulations and experiments on real quantum hardware are discussed on Section \ref{sec:Results}. A study of the efficiency of the generated oracle is provided in Section \ref{sec:efficiency}. Finally, the conclusions and future work are presented in Section \ref{sec:conclusions}.

\section{Background}
\label{sec:background}
The oracle has been identified as a pattern for quantum algorithms \cite{Leymann-QuantumAlgorithms}. An oracle can be thought as a black box performing a function that is used as an input by another algorithm \cite{oracles_as_black_boxes}. 
Thus, how an Oracle works is not a matter of concern for the algorithm that uses it. These features make oracles a good tool for quantum software reuse. 
Apart of Grover's, many other algorithms employ oracles. Some well known examples are Deutsch-Jozsa \cite{Deutsch1992RapidSO}, Simon \cite{Simon_1997} or Bernstein-Vazirani \cite{Bernstein_Vazirani}. 

There are two main types of oracles described in the literature \cite{typesOracles2019}, probability oracles and phase oracles. On one hand, probability oracles are common in quantum optimization procedures. On the other hand, phase oracles are used in quantum algorithms (such as Grover's) and encode a function in the phase of the quantum states. In the case of Grover's algorithm, the function implemented by the oracle recognises the desired states. As mentioned above, the less-than oracle implemented in this work is a phase oracle. 



Thinking of oracles in terms of black boxes favours the reusability of quantum software. One would then expect any reusable software to have the best possible quality attributes. Today's quantum devices are subject to decoherence. Thus, the depth of the executing circuits is a key aspect to maximise their reliability. The greater the depth of a circuit, the more it is exposed to decoherence and the lower its reliability \cite{depth-decoherence}. Therefore, if the depth of the circuit that implements a quantum algorithm is not taken into account, it can end up producing a result that is indistinguishable from pure noise \cite{Preskill2018quantumcomputingin}. So keeping depth of oracles optimised, contributes to optimise their quality attributes and their chance to be reused.

 


In order to properly contextualise the depth of the circuit associated to the oracle proposed in this paper, we compare it to \textit{UnitaryGate}, the generation method implemented in Qiskit \cite{qiskit}. This method is based on \cite{Isometries, Cross_2019}. There are other automatic methods for oracle generation, such as \cite{automaticGeneration2023}, however, as Qiskit is one of the most used quantum SDKs, we have chosen to compare the performance of the less-than implementation with \textit{UnitaryGate} generated circuit, as also does \cite{multicontrol2022}. In addition, our method avoids the use of ancilla qubits.

In this paper not only the concept of the oracle is provided, also a classical algorithm is detailed to automatically generate an efficient implementation of the proposed less-than oracle. This algorithm could be included as a pattern for the less-than problem in intelligent code generators such as the one mentioned in \cite{code-generators}. This implementation improves the one provided by \textit{UnitaryGate} in terms of depth, the achieved improvement is enough to outperform it in real quantum hardware. 

\section{Algorithm for the automatic generation of the oracle} \label{sec:classical}

In this section we detail the classical algorithm designed to automate the building of the oracle circuit. This oracle will mark with a $\pi$-phase all quantum states which represent natural numbers strictly smaller than a given one.  The oracle is represented by a unitary matrix of the form:
 

    

\begin{equation}
\begin{pmatrix}\label{matrix}
-1 &  &  &  &  &  \\
 & \ddots &  &  &  &  \\
 &  & -1 & \multicolumn{3}{c}{\raisebox{\dimexpr\normalbaselineskip+.7\ht\strutbox-.5\height}[0pt][0pt]{\scalebox{2}{$0$}}} \\
 &  &  & 1 &  &  \\
 &  &  &  & \ddots & \\
 \multicolumn{3}{c}{\raisebox{\dimexpr\normalbaselineskip+.7\ht\strutbox-.5\height}[0pt][0pt]{\scalebox{2}{$0$}}} &  &  & 1  \\
\end{pmatrix} 
\end{equation}

The classical algorithm we propose is based on the idea that, in order to compare the binary representation of two natural numbers $n_1$ and $n_2$, it is needed to look for the first bit  (starting at the most significant one) which differs in these two numbers. Once this bit is found, it is compared for the two numbers, the one whose bit is $0$ is the smaller one. E.g. when comparing\footnote{In numbers like $101_2$ the 2 in the subindex means that it is a binary number.} $n_1=111_2 = 7$ and $n_2=101_2=5$, the most significant bits are both equal to 1, the second most significant bits differ. The bit which is 0 belongs to $n_2$, hence $n_2$ is the smaller one.

The pseudocode of this algorithm is detailed in \ref{algorithm}. It follows an explanation which describes it in detail.
\begin{algorithm}[h]
\label{algorithm}
\SetAlgoLined
 
 \SetKwProg{Init}{init}{}{}
 \KwIn{Number of qubits $n$ and a natural number $m$ where $0<m<2^n$}
 \KwOut{Quantum Circuit which gives a $\pi$-phase   to all states representing binary forms of natural numbers smaller than $m$}
  $m_{\text{binary}} \gets \text{binary}(m)$\;
  $QuantumCircuit \gets \text{Circuit of $n$ qubits}$\;
  $b_{n-1} \gets 1$st bit (most significant) of $m_{\text{binary}}$\;
 \eIf{$b_{n-1}=0$}{
   $X$ gate to $q_{n-1}$\;
 }{
   $X$ gate to $q_{n-1}$\;
   $Z$ gate to $q_{n-1}$\;
   $X$ gate to $q_{n-1}$\;
 }
 \For{$i\gets n-2$ \KwTo $0$}{
  $b_i \gets i$-th bit of $m_{\text{binary}}$\;
  $q_i \gets i$-th qubit of $QuantumCircuit$\;
  \eIf{$b_i=0$}{
   $X$ gate to $q_i$\;
   }{
   $X$ gate to $q_i$\;
   $CZ^{\otimes (n-i)}$ gate to qubits $q_{n-1},\,\ldots,\,q_i$\;
   $X$ gate to $q_i$\;
  }
 }
 \For {$i \gets 0$ to $n-1$}{
 \If{$b_i = 0$}{
 $X$ gate to $q_i$\;
 }
 }
 \caption{Less-than oracle classical builder}
\end{algorithm}

The algorithm needs as inputs the number of qubits $n$ and a natural number $m$ where $0<m<2^n$. As stated before, the output of our algorithm produces a quantum circuit with $n$ qubits which gives a $\pi$-phase to all states representing natural numbers strictly smaller than $m$.

For the algorithm to work as intended is needed to initialise all qubits to a uniform superposition of 0s and 1s by applying a Hadamard gate to each one.

The first step of the algorithm is converting the number $m$ to binary form, $m_\bi = b_{n-1}\ldots b_0$, with $n$ bits\footnote{In the case $n=4$ and $m=3$, the binary form would be $m_\bi=0011.$}. This is always possible because $m<2^n$. 

After that, it proceeds to check the first bit (most significant) of $m_{\text{binary}}$, $b_{n-1}$. If it is 1, hence $m\geq 2^{n-1}$, then the gates $XZX$ are applied to the qubit $q_{n-1}$ of the quantum circuit (\ref{eq:XZXq1}). This gives a $\pi$-phase to the states of the form $|0\,q_{n-2}\ldots q_0\rangle$, which represent numbers smaller than $2^{n-1}$, thus smaller than $m$.

\begin{align}\label{eq:XZXq1}
\begin{split}
&XZX \otimes I^{\otimes (n-1)} |q_{n-1}\ldots q_0\rangle = \\
= - &\dfrac{1}{\sqrt{2}} |0\,q_{n-2}\ldots q_0\rangle + \dfrac{1}{\sqrt{2}} |1\,q_{n-2}\ldots q_0\rangle  
\end{split}  
\end{align}
where $I^{\otimes (n-1)}$ is the identity gate applied to $n-1$ qubits, in this case the least significant ones.

Else, if $b_{n-1}$ is 0, an $X$ gate is applied to $q_{n-1}$, which will be reversed at the end of the circuit.

The next step is general for the rest of the positions, $i\in\{n-2,\ldots,0\}$. 
If $b_i$ is 0, an X gate is applied to $q_i$. In case $b_i$ is 1, the gates
\begin{equation}
   I^{\otimes (n-i-1)} \otimes X \cdot CZ^{\otimes (n-i)} \cdot I^{\otimes (n-i-1)} \otimes X 
\end{equation}
are applied to $\{q_{n-1},\ldots,q_i\}$. This gives a $\pi$-phase to the states $|b_{n-1}\ldots b_{i+1}\,0 \, q_{i-1}\ldots q_0\rangle$. As explained before, if the $i$-th bit of $m_\bi$ is 1, all numbers with 0 in the $i$-th bit and the same $n-i-1$ first bits as $m_\bi$ ($b_{n-1}\ldots b_{i+1}$) are smaller than $m$.

Finally, an $X$ gate is applied to each $q_i$ such that $b_i=0$, 
\begin{equation}\label{eq:X_final_0s}
\prod_{i=0\,:\,b_i=0}^{n-1} I^{\otimes (n-i-1)} \otimes X \otimes I^{\otimes (i)}    
\end{equation}
This is done because another $X$ gate was applied to them in previous steps of the algorithm, so operation (\ref{eq:X_final_0s}) returns these qubits to the initial state, except for a possible $\pi$-phase.

\begin{figure}[h]
\centering
\includegraphics{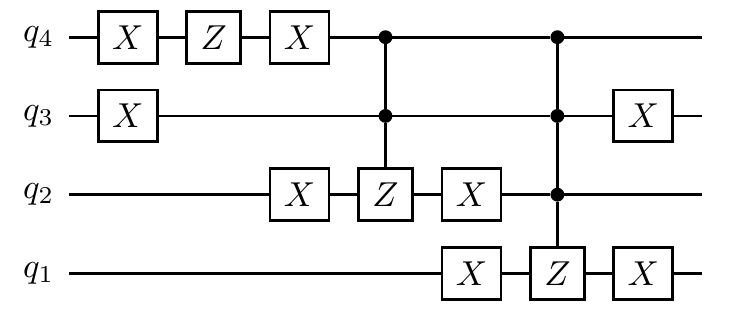}
\caption{Oracle generated taking $m=11 = 1011_2$ with 4 qubits.}
\label{fig:oracle-less-than-11}
\end{figure}

As an illustrative example of the algorithm output, we show the oracle generated taking $m=11=1011_2$ in figure \ref{fig:oracle-less-than-11}.

From this point, the implementation of the greater-than oracle mentioned above is trivial. Even more, the oracles greater-than and less-than can be composed in a way that, by combining them, we obtain ranges of integers. These oracles also satisfy the commutative and associative properties. Although this aspect is not exploited in this work it is one of the key motivations behind it. This combinations and further examples of less-than oracles may be found in the repository included in \hyperref[sec:code-data]{Code and Data}.

\section{Results}\label{sec:Results}
The best way to show the functionality of the less-than oracle implementation presented in this work is through simulations of different examples of the generated circuits and experiments on real quantum devices\footnote{The code for the automatic generation and the data of the simulations and experiments can be found in \hyperref[sec:code-data]{Code and data}.}. For this paper, we have used Qiskit \cite{qiskit} for generating the quantum circuits and running the experiments. 

\begin{figure}[h]
\centering
\includegraphics[width=0.47\textwidth]{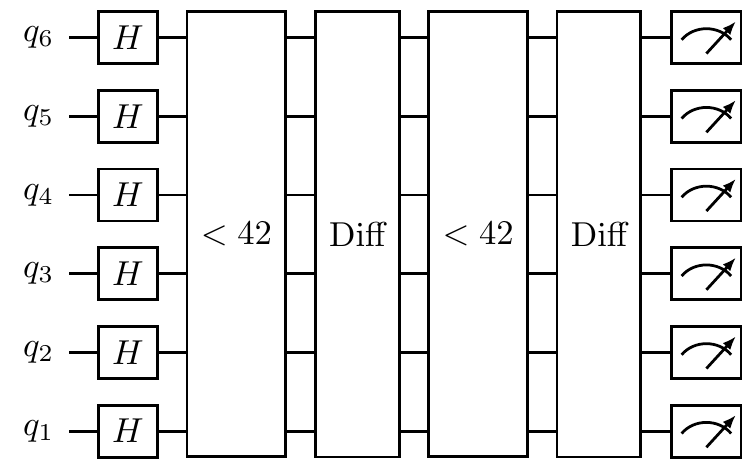}
\caption{Full circuit implementing operation less-than 42 with 6 qubits.}\label{fig:full-circuit-less-than-42}
\end{figure}

We show the results of amplifying the desired states as described by chapter 6 of \cite{nielsen00}. As stated before, to realise that Amplitude Amplification a diffusion operator is applied after the oracle. Depending on the fraction $N/M$ (being $M$ the number of desired states and $N$ the total number of states), we need to repeat the pair oracle-diffuser several times, with a maximum of $\lceil \pi/4\sqrt{N/M}\rceil$ times. In figure \ref{fig:full-circuit-less-than-42}, we show an example of the full circuit implementing the operation less-than 42 with 6 qubits.

All our results are generated by running the circuit several times, each of those runs is called shot. Each shot is the result provided by measuring after running the circuit. Hence, the output of one shot is just one state. Figures in this section represent the relative frequency (y-axis) of each state (x-axis) measured in the several shots realised. The results shown are generated by running 20,000 shots of each circuit. This choice is based on the maximum number of shots allowed in IBM real quantum devices. This is done to properly compare simulations and real experiments.

\subsection{Simulations}
Simulations shown in this section were run on a standard laptop\footnote{CPU: Intel(R) Core(TM) i5-7200U CPU @ 2.50GHz, 2701 Mhz. RAM: 8GM.} and execution time ranged between 8 and 10 seconds for each of them.

\begin{figure*}
    \centering
    \includegraphics[width=0.9\textwidth]{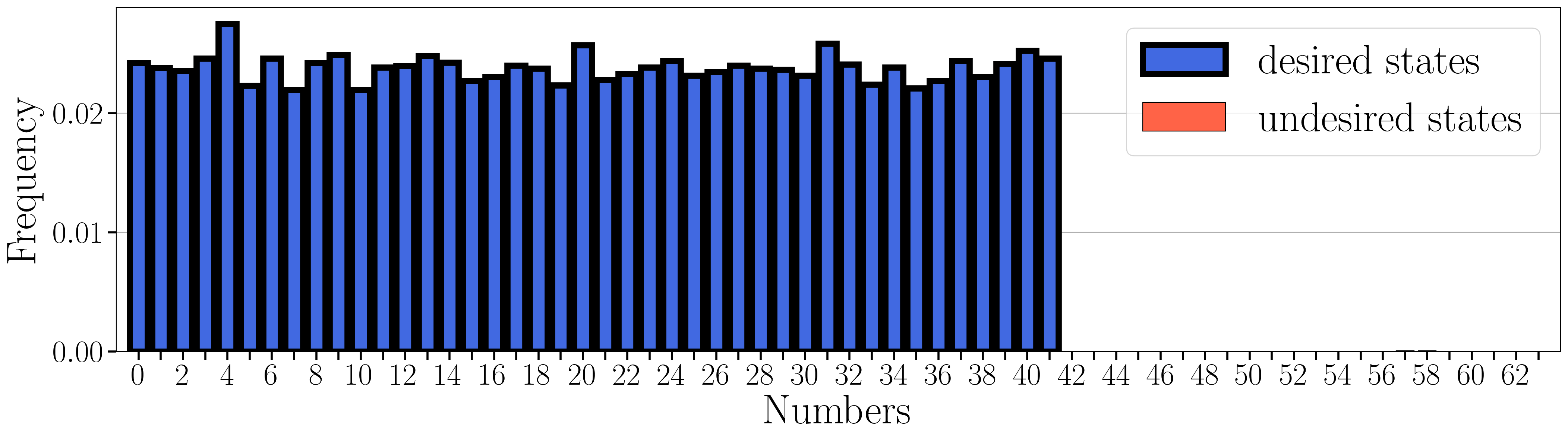}
        \caption{Simulation of less-than 42 amplification with 6 qubits and 20,000 shots.}
    \label{fig:less-than-42}
\end{figure*}

\begin{figure*}
    \centering
    \includegraphics[width=0.9\textwidth]{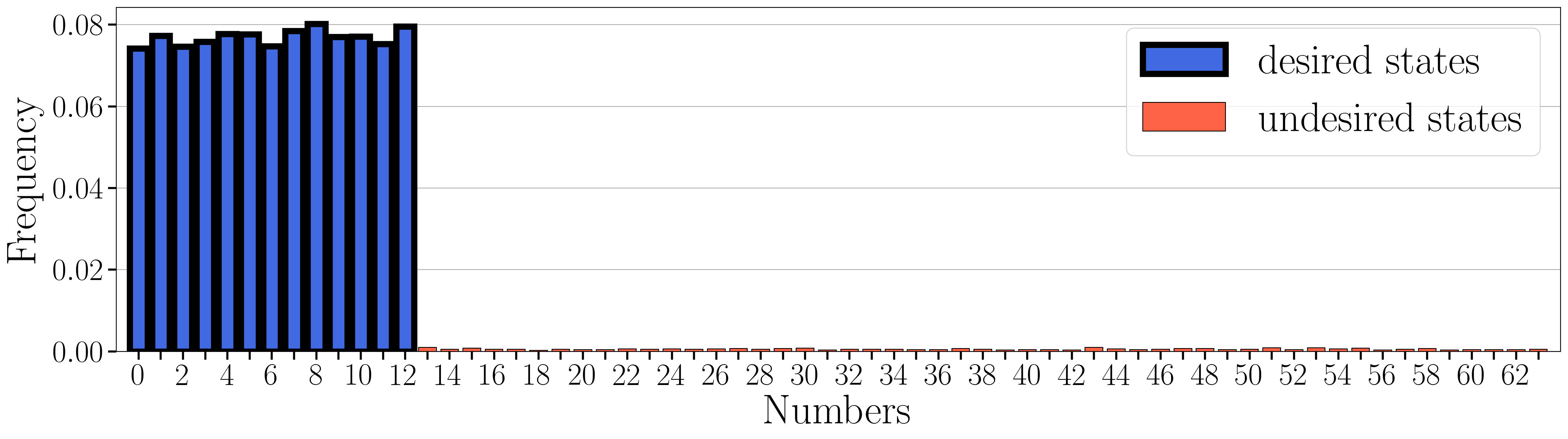}
        \caption{Simulation of less-than 13 amplification with 6 qubits and 20,000 shots.}
    \label{fig:less-than-13}
\end{figure*}

Figures \ref{fig:less-than-42} and \ref{fig:less-than-13} are examples with 6-qubit circuits. Figure \ref{fig:less-than-42} shows the result of the less-than 42 amplification. The desired state is a superposition of the first 42 states:
\begin{equation}
\dfrac{1}{\sqrt{42}}\sum_{i=0}^{41} |i\rangle    
\end{equation}

The pair oracle-diffuser had to be repeated 2 times to reach maximum amplification of the desired state. Full circuit can be seen in Figure \ref{fig:full-circuit-less-than-42}. It can be seen that the results are exactly as we would expect save for minor frequency differences. This is due to the number of shots simulated. The frequency of each number tends to the theoretical value (\textcolor{blue}{$1/42$} in this case) when the number of shots tends to infinity.

Figure \ref{fig:less-than-13} displays results of the less-than 13 operation. In contrast with results in figure \ref{fig:less-than-42}, in this case there are some small occurrences of non-desired states. The cause of this is that the best amplification possible, which occurs with 1 iteration, is not exactly 100\%.

\begin{figure*}
    \centering
    \includegraphics[width=0.9\textwidth]{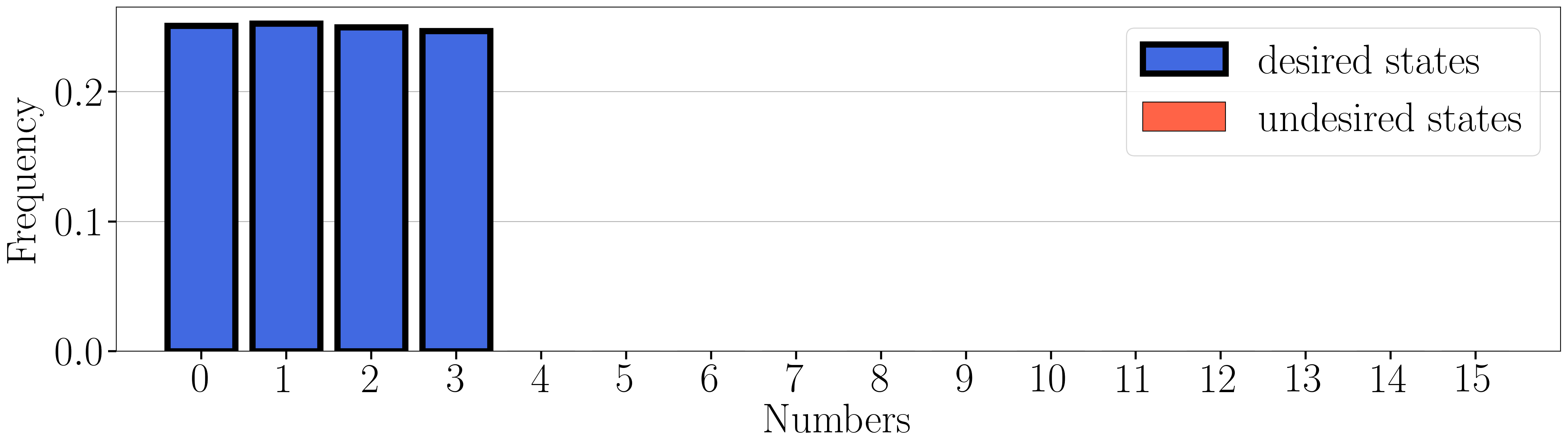}
        \caption{Simulation of less-than 4 amplification with 4 qubits and 20,000 shots.}
    \label{fig:less-than-4}
\end{figure*}

Results from less-than 4 operation on a 4-qubit circuit are shown in figure \ref{fig:less-than-4}. In this case, the number of iterations of the pair oracle-diffuser is also 1, however the amplification reaches 100\% measurement probability of the desired states. The main objective of this example is to compare a simulation with an execution on a real quantum device. Such results are available in the next section. 

The numbers used to perform the less-than operations in this section (42, 13 and 4) have been chosen to showcase that Amplitude Amplification, even in absence of noise, does not always perfectly amplify the desired states.

\subsection{Experiments on real quantum hardware}
\begin{figure*}
    \centering
    \includegraphics[width=0.9\textwidth]{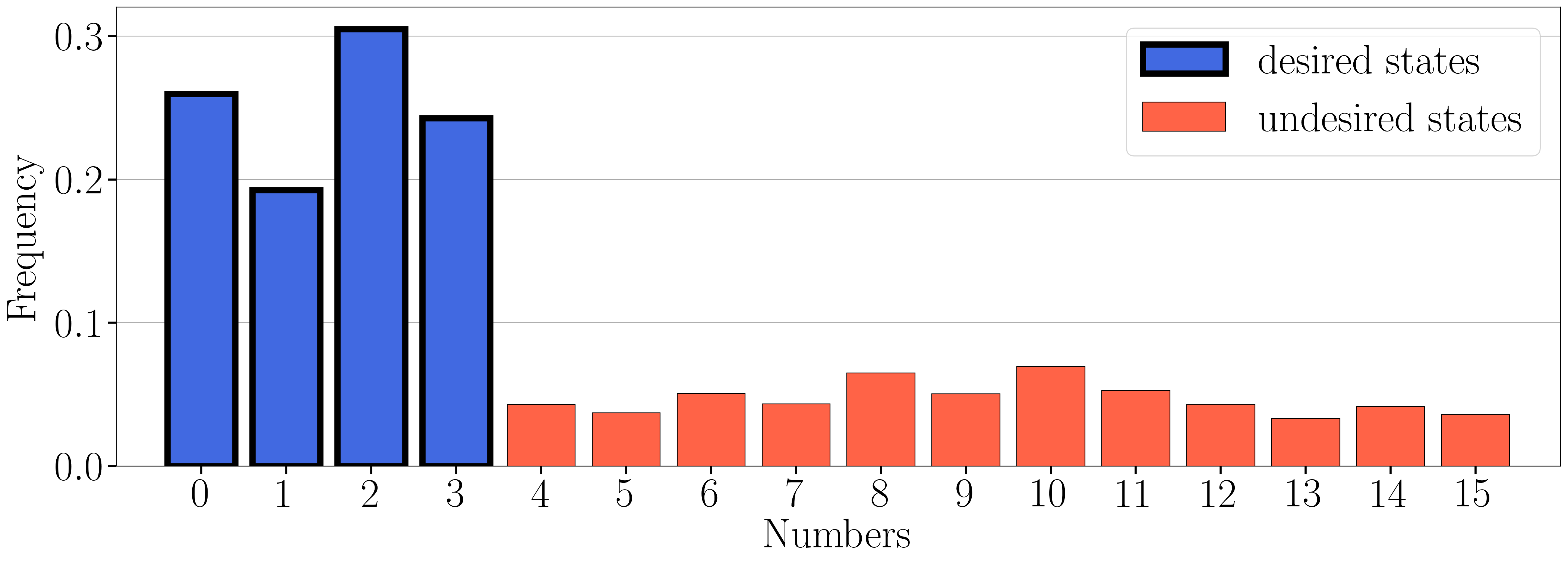}
        \caption{Real experiment of less-than 4 amplification with 4 qubits and 20,000 shots.}
    \label{fig:less-than-4-real-DJ}
\end{figure*}

We have run the last simulation, less-than 4 amplification on a 4-qubit circuit, on a real quantum device with 20,000 shots as well. We have chosen this operation specifically because for 4 qubits the full amplification circuit is the one with lower depth. Results of the experiment can be seen in figure \ref{fig:less-than-4-real-DJ}. The error is noticeable, which is an expected behaviour with real quantum hardware.

Despite these errors, the desired states have a combined frequency of $\approx 63\%$, even though they are 4 states out of 16 in total, $25\%$. Hence we managed to amplify the measurement probability of the desired states by a $\approx 2.5$ factor on an actual quantum device.

The device used for this experiment was an IBM machine, \textit{ibm\_nairobi}, on November 18th, 2022 13:22 UTC. This device uses a Falcon r5.11H processor. The calibration data of the device at the time of the experiment can be found in the repository in \hyperref[sec:code-data]{Code and data}.

\section{Efficiency of the generated oracle} \label{sec:efficiency}

In this section we focus on two main aspects related with efficiency: number of qubits and depth.

The number of qubits is crucial given that in current quantum devices the number of available qubits is very limited  \cite{quantumhardware}. Depth (the number of consecutive gates) represents the physical implementation time for the algorithm \cite{depthoptimization}. On actual quantum hardware only low-depth circuits are realizable, otherwise the noise dominates the result \cite{Preskill2018quantumcomputingin}.

\subsection{Qubits}
The method we propose does not use any ancilla qubits. The only required qubits are the ones needed for the full superposition of $N$ states, being $N=2^{n}$, where $n$ is the number of qubits. Hence, our method is as efficient as possible in terms of number of qubits.


\subsection{Depth}
To give context to the depth of our method, we have chosen to compare the depth with the oracle circuit automatically generated by Qiskit method \textit{UnitaryGate}. The \textit{UnitaryGate} method needs as input a unitary matrix. In this case, the matrix we use as input is the one stated in matrix (\ref{matrix}).

In order to do the comparison we generate the same circuit with both methods for each number $k$ with $0<k<N$.

To properly compare the two methods, we have
transpiled all circuits to one of the IBM quantum computers. The backend used was \textit{fake\_washington\_v2} which has the same properties (gate set, coupling map, etc.) as the real one.

\begin{figure*}
    \centering
    \includegraphics[width=0.9\textwidth]{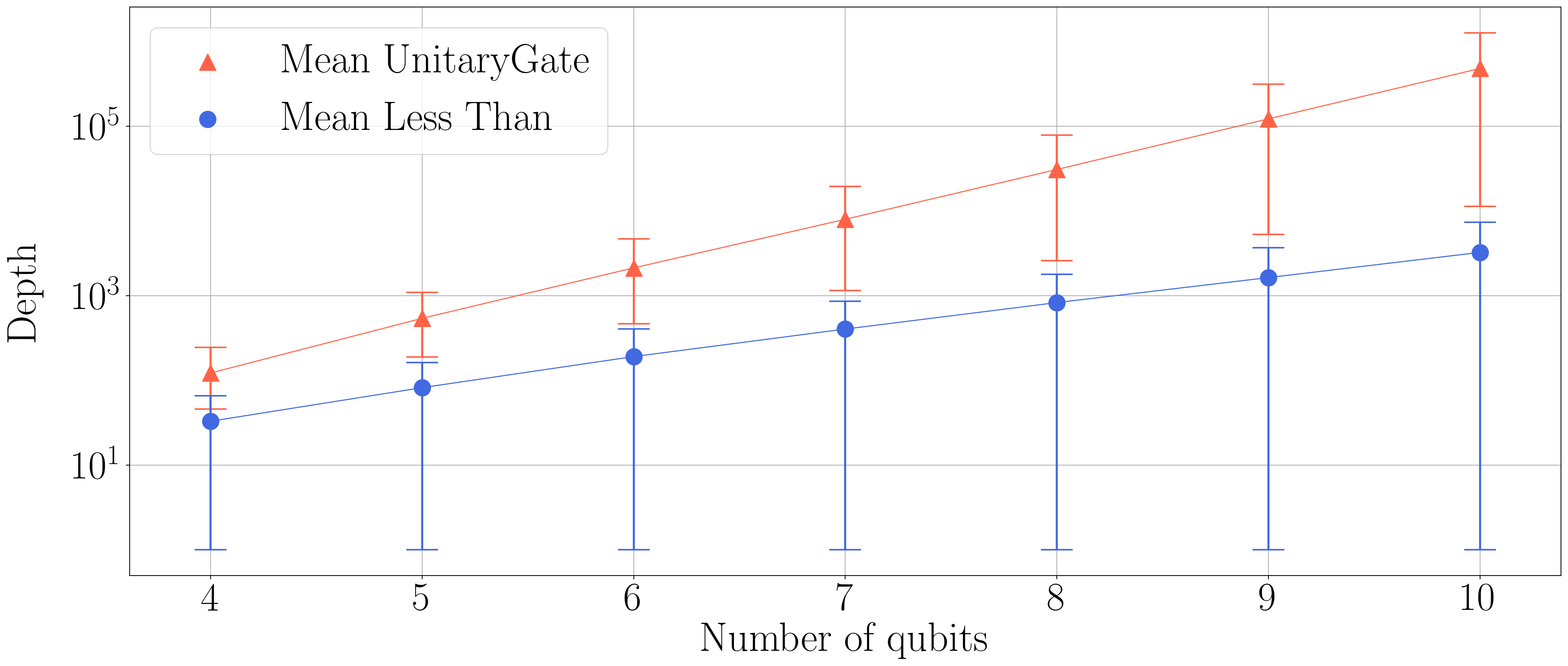}
    \centering
    \caption{Comparison of depth using UnitaryGate method vs Less than. Lines between points are visual guides and do not represent any data.}
    \label{fig:depth-comparison}
\end{figure*}

The results of the evaluations are shown on Figure \ref{fig:depth-comparison}. The dots and triangles are the mean depth of the less-than oracle circuits and the UnitaryGate generated ones, respectively. The vertical bars represent the maximum and minimum depth for that number of qubits. The scale is logarithmic.  While for 4 qubits the minimum depth of UnitaryGate is lower than the maximum depth of our method, this happens for circuits with different $k$. We have checked that the depth of the circuits generated by the algorithm we propose is always lower than their counterparts generated by \textit{UnitaryGate}.

\begin{figure*}
    \centering
    \includegraphics[width=0.9\textwidth]{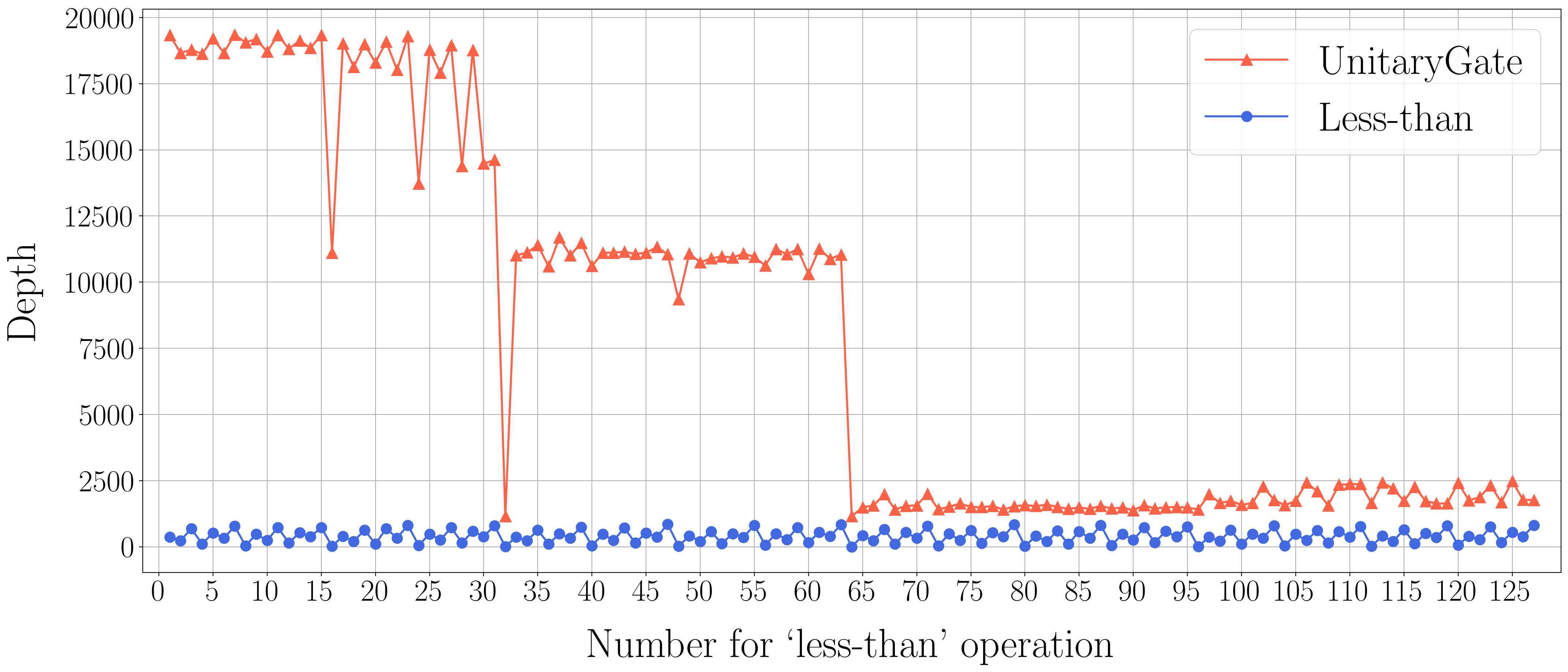}
    \centering
    \caption{ Depth of oracle less-than on 7 qubit circuit. Lines between points are just visual guides, they do not represent any data.}
    \label{fig:depth-7-qubits}
\end{figure*}

The depth for every circuit with 7 qubits can be seen on Figure \ref{fig:depth-7-qubits}. The same plot for the remainder numbers of qubits can be found on section \ref{sec:code-data}.

The main result from this analysis is that for each number our method produces a circuit with lower depth than Qiskit method \textit{UnitaryGate} and that difference grows with the number of qubits. The largest differences occur for low numbers ($<N/2$) with a high number of qubits. It can be noted that oracle circuits for numbers which are powers of 2 have reduced depths with both methods. Even in these cases, our method is a great improvement from Qiskit \textit{UnitaryGate}. For instance, our oracle implementation for the less-than 32  operation with $n=7$ qubits has a depth of 11. The one generated by \textit{UnitaryGate} has a depth of 1625.

\begin{figure*}
    \centering
    \includegraphics[width=0.9\textwidth]{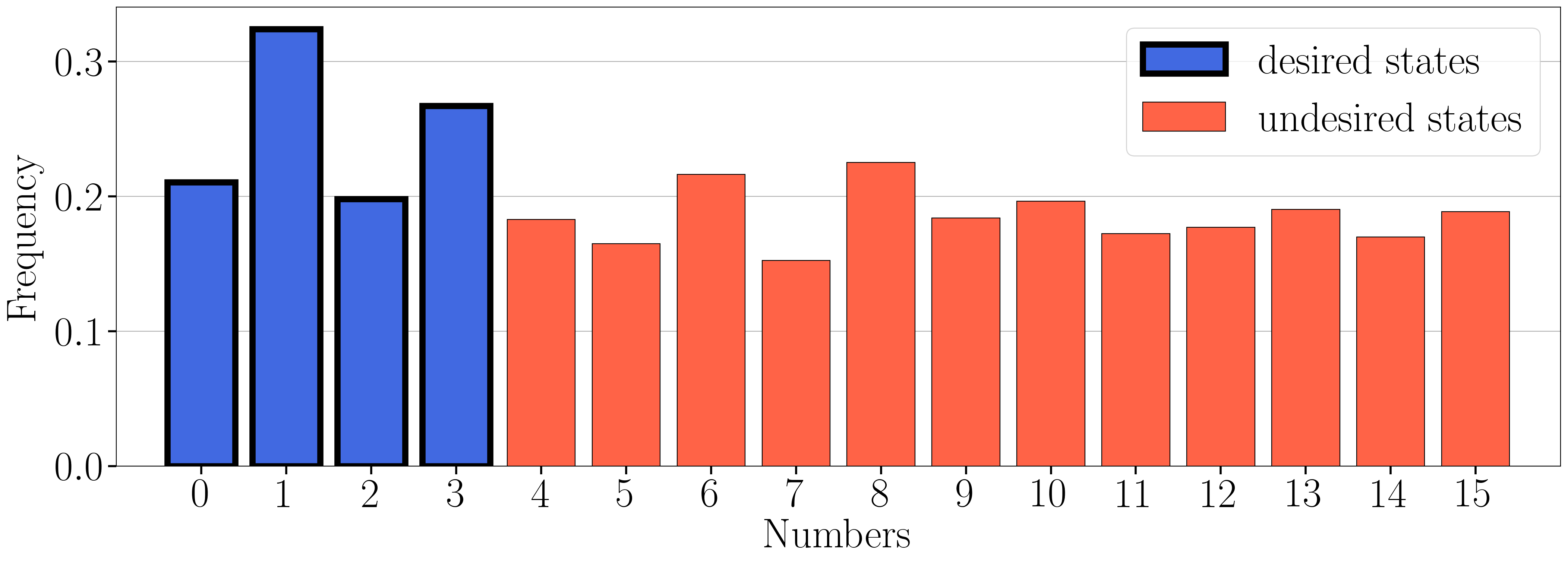}
        \caption{Real experiment of less-than 4 amplification with Qiskit \textit{UnitaryGate} method with 4 qubits and 20,000 shots.}
    \label{fig:less-than-4-real-unitary}
\end{figure*}

Next, we show that \textit{UnitaryGate} generated oracle circuit fails at being usable on real quantum hardware in cases where our method is successful. We have conducted again the experiment displayed at figure \ref{fig:less-than-4-real-DJ} but using the \textit{UnitaryGate} instead of our method to generate the circuit. This experiment was executed on the IBM machine \textit{ibm\_nairobi}, on November 18th, 2022 13:00 UTC. The device had the same calibration for both experiments. The results of the experiment are shown on figure \ref{fig:less-than-4-real-unitary}.

In the experiment carried out with our method, the desired states had a combined frequency of $\approx 63\%$. These states are 4 states out of 16 in total, $25\%$. On the other hand, in the new experiment with the method \textit{UnitaryGate}, these states have a combined frequency of $\approx 31\%$. Our method increased the probability of the desired states by a $\approx 2.5$ factor, while \textit{UnitaryGate} method increased it by a $\approx 1.2$ factor. As these two experiments were conducted using the same hardware with the same calibration, the only difference is the depth of the circuits. Therefore, we believe depth is the reason our method provides a much more satisfactory result than \textit{UnitaryGate}.

\section{Conclusions and future work} \label{sec:conclusions}
We have presented a classical algorithm capable of building a phase-marking oracle circuit that performs a less-than operation. To exemplify its functionality, some experiments were conducted on both simulations and real quantum hardware. We have made a study on efficiency, focusing on number of qubits and depth. We have shown that no ancilla qubits are required. It has also been shown that the depth of this oracle implementation is always lower than the one generated by Qiskit automatic methods. This depth difference is significant enough for our method to outperform the Qiskit method on real quantum hardware.

As mentioned above, the oracle proposed in this work is the first step in the creation of a bigger set of oracles including ones like greater-than or range of integers. By combining these and other, similar, oracles we expect to be able to create a set of efficient tools for working with integers that can be reused as black box operations. These tools can then be composed by quantum software developers in the creation of complex algorithms.

\section*{Acknowledgments}
This work has been financially supported by the Ministry of Economic Affairs and Digital Transformation of the Spanish Government through the QUANTUM ENIA project call - Quantum Spain project, by the Spanish Ministry of Science and Innovation under project PID2021-124054OB-C31, by the Regional Ministry of Economy, Science and Digital Agenda, and the Department of Economy and Infrastructure of the Government of Extremadura under project GR21133, and by the European Union through the Recovery, Transformation and Resilience Plan - NextGenerationEU within the framework of the Digital Spain 2026 Agenda.

We acknowledge the use of IBM Quantum services for this work. The views expressed are those of the authors, and do not reflect the official policy or position of IBM or the IBM Quantum team.

We are grateful to COMPUTAEX Foundation for allowing us to use the supercomputing facilities (LUSITANIA II) for calculations.

\section*{ Code and data}\label{sec:code-data}
Data and code can be found in
\href{https://github.com/JSRivero/Less-than-oracle}{https://github.com/JSRivero/Less-than-oracle}.
\bibliographystyle{IEEEtran}
\bibliography{IEEEabrv,bibliography}

\end{document}